\documentstyle[12pt,twoside,fleqn,espcrc1]{article}

\def\nn{\nonumber}            
\def\beq{\begin{eqnarray}}    
\def\enq{\end{eqnarray}}      
\def\ap{\left.}               
\def\at{\left(}               
\def\aq{\left[}               
\def\cp{\right.}              
\def\ct{\right)}              
\def\cq{\right]}              
\def\R{{\hbox{{\rm I}\kern-.2em\hbox{\rm R}}}}   
\def\H{{\hbox{{\rm I}\kern-.2em\hbox{\rm H}}}}   
\def\N{{\hbox{{\rm I}\kern-.2em\hbox{\rm N}}}}   
\def\C{{\ \hbox{{\rm I}\kern-.6em\hbox{\bf C}}}} 
\def\Z{{\hbox{{\rm Z}\kern-.4em\hbox{\rm Z}}}}   
\newcommand{\fr}[2]{\mbox{$\frac{#1}{#2}$}}      

\def\be{\beta}

\newcommand{\detau}{\partial_{\tau}}
\newcommand{\detaud}{\partial_{\tau}^{2}}

\newcommand{\nablad}{\nabla^{2}}

\newcommand{\matrice}[4]{\left(\begin{array}{cc} {#1} & {#2} \\ {#3} & {#4} \end{array} \right) }

\newcommand{\AmS}{{\protect\the\textfont2
  A\kern-.1667em\lower.5ex\hbox{M}\kern-.125emS}}

\title{Multiplicative anomaly and finite charge density}

\author{Antonio Filippi\address{Theoretical Physics Group, Imperial
   College, \\Prince Consort Road, London SW7 2BZ, United Kingdom}
        \thanks{Talk given at the Workshop on QCD at Finite Baryon
Density, April 27-30, 1998, Bielefeld, Germany. The author wishes to
acknowledge financial support from the European Commission 
under TMR contract N. ERBFMBICT972020 and, previously, 
from the Foundation Blanceflor Boncompagni-Ludovisi, n\'ee Bildt.}}

\begin{document}
\maketitle

\begin{abstract}
When dealing with zeta-function regularized functional
determinants of matrix valued differential operators, an
additional term, overlooked until now and due to the
multiplicative anomaly, may arise. The presence and physical relevance
of this term is discussed in the case of a charged bosonic field at finite charge density and other possible applications are mentioned.
\end{abstract}

\section{INTRODUCTION}

In field theory we often have to deal with functional determinants of
differential operators. These, as formal products of infinite eigenvalues, are
divergent objects (UV divergence) and a regularization scheme
is therefore necessary.
One of the most successful and powerful ones is the zeta-function
regularization method \cite{raysin,dowcri,haw,bcvz,libro}. 
It permits us to give a meaning to the ill defined quantity
$\ln\det A $, where $A$ is a second order elliptic differential
operator, through the zeta
function  $\zeta(s|A)=\mbox{Tr}\, A^{-s}$,
which is well defined for a sufficiently large real part of $s$ and can be analytically
continued to a function meromorphic in all the plane and analytic at
$s=0$. As such its derivative to respect to $s$ at zero is well defined and
the logarithm of the zeta-function regularized functional determinant
will then be defined by
\begin{eqnarray} 
\ln\det\frac{A}{M^2} =-\zeta'(0|A)-\zeta(0|A)\ln M^2 ,
\end{eqnarray}
where $M^2$ is a renormalization scale mass.

Sometimes, however, the differential operator takes a matrix form in
the field space, as is the case with the two real components
$\phi_i$ of a complex scalar field. 
In this case we end up evaluating a quantity of the
form $\ln \det (A B)$, with $A$ and $B$ two commuting pseudo-differential operators.
The fact is that it is not always true that the equality $\ln \det
(AB)=\ln \det(A)+\ln \det (B)$ holds.
On the contrary, an additional term $a(A,B)$,
called the multiplicative anomaly \cite{wod,konvis,kas},
may be present on the right
hand-side and eventually have physical relevance \cite{evz,efvz}.
In this work I will introduce this quantity, compute it and
analyse its physical relevance in the case of a charged scalar field
at finite temperature and charge density, as well as present other
possible physical systems in which it could play a role.  

This work has been developed in collaboration with E. Elizalde, in
Barcelona (Spain), and L. Vanzo and S. Zerbini, in Trento (Italy). My
thank goes also to R. Rivers and T. Evans for stimulating discussions.

\section{THE BOSE GAS AT FINITE CHARGE DENSITY}

The relativistic complex scalar field at finite temperature in the 
presence of a net charge density has given rise to a certain interest
during recent years \cite{kap,habwel,bbd,bd,kirtom,chemical}.

In our recent paper \cite{efvz} it is shown that, in a coherent regularized
approach, the multiplicative anomaly, overlooked until now, could
play a role in this system. I will try here to outline the results
avoiding the mathematical machinery. For clarity, I will mainly restrict
myself to four space-time dimensions although the system has
been studied in generic $D$ dimensions.

The relevant quantity for my proposes is the grand canonical partition
function, which, for this system, is \cite{kap,habwel,chemical}
\begin{eqnarray}
Z_{\beta}(\mu)=\mbox{Tr}\,   e^{-\beta (H-\mu Q)} =
\int_{\phi(\tau)=\phi(\tau+\beta)}[d \phi_i]
e^{-\frac{1}{2}\int_0^\beta d\tau \int d^3x \phi_iA_{ij}\phi_j}
\:,\label{dddd}
\end{eqnarray} 
where $H$ is the Hamiltonian of the system, $Q$ the charge and $\beta$
the inverse of the temperature. 
$A_{ij}$ is the elliptic, non-self-adjoint, matrix valued,
differential operator
\beq
\matrice{-\detaud-\nablad+m^{2}-e^2\mu^{2}}{-2ie\mu\detau}{2ie\mu\detau}{-\detaud-\nablad+m^{2}-e^2\mu^{2}}.
\enq

In this case computing the partition function requires taking both an algebraic
determinant and a functional one. The standard procedure consists
in taking the algebraic one first \cite{kap,bbd,bd,kirtom}. Stimulated by
a recent criticism \cite{dowker}, we showed the validity 
of this procedure \cite{rdowker}.

Now, we have two possible factorizations for this
algebraic determinant:
\beq
\ln Z_{\be}(\mu)=-\frac{1}{2}\ln\det \left\| \frac{A_{ij}}{M^2}\right\| =-\frac{1}{2}\ln\det \aq
\frac{L_+}{M^2}\frac{L_-}{M^2}\cq =-\frac{1}{2}\ln\det \aq
\frac{K_+}{M^2}\frac{K_-}{M^2}\cq
\label{o2}\:,
\enq
where:
\beq
K_{\pm}=-\nablad+m^{2}+\left(i\detau\pm ie\mu\right)^2 \ \ \ \ \ \ \ \
\ \  L_{\pm}=-\detaud+\left(\sqrt{-\nablad+m^{2}}\pm e\mu\right)^2 .
\enq

I will avoid here the standard steps that lead to the computation of
the logarithm of the partition function, as the reader will find them in
greater detail in ref. \cite{efvz}.

Assuming, as in the precedent literature, the validity of the identity
$\ln \det(AB)=\ln \det(A)+\ln \det (B)$ and disregarding the anomaly,
we obtain, for the $K_{\pm}$ factorization,
\begin{eqnarray}
\!\!\!\!\!\ln Z_{\be}(K_+,K_-)\!\!\!\!&=&\!\!\!\!\frac{\be V }{32\pi^2}\aq
m^4 (\ln \fr{m^2}{M^2}-3/2) \cq \nn \\
&-&\!\!\!\! V\!\! \int\!\frac{d^{3}k}{(2\pi)^{3}}\!
\ln(1\!-\!e^{-\be(\sqrt{k^{2}+m^{2}}-e\mu)})\! -\! V\!\! \int\!\frac{d^{3}k}{(2\pi)^{3}}\! \ln(1\!-\!e^{-\be(\sqrt{k^{2}+m^{2}}+e\mu)}) ,
\label{441}
\enq
where the expected contributions for vacuum, particles and
antiparticles are manifest. {}From now on I will
represent the thermal contributions as $S(\be,\mu)$.

Similar manipulations can be done for the other factorization
$L_{\pm}$. In this case though,  the chemical potential does not
appear in the sum over Matsubara frequencies, but remains with the
momentum integral and therefore the term linear in $\beta$ will be chemical
potential dependent:
\beq
\ln Z_{\be}(L_+,L_-)&=&\frac{\be V }{32\pi^2}\aq
m^4 (\ln \fr{m^2}{M^2}-3/2) \cq+ \frac{\be V }{8\pi^2} \at
\frac{e^4\mu^4}{3}-e^2\mu^2 m^2 \ct + S(\be,\mu)\: .
\label{44}
\enq
In this system the importance of the multiplicative anomaly is
therefore manifest. Despite having $ \ln (K_{-}K_{+}) =\ln (L_{-}L_{+})$, 
these two options give two different results for a zeta-function regularized partition function {\em if the multiplicative
anomaly is disregarded}.

\section{THE MULTIPLICATIVE ANOMALY}

The multiplicative anomaly \cite{wod,konvis,kas} is defined as
\begin{eqnarray} 
a_D(A,B)=\ln \det (AB)-\ln \det(A)-\ln \det (B) 
\end{eqnarray}
where the determinants of the two elliptic operators are defined by
means of the zeta-function method. I recall that $D$ are the space-time
dimensions.
In principle, it could be computed directly as difference of the
involved quantities. In reality, actual calculations are very
complicated even for simpler operators. We can fortunately resort to
Wodzicki's results for a remarkably neat recipe. 

For any classical pseudo-differential operator $A$ there exists
 a complete symbol $A(x,k)=e^{-ikx}Ae^{ikx}$.
This admits an asymptotic expansion for $|k| \to \infty$,
\begin{eqnarray} 
A(x,k)\sim\sum_{j=0} A_{a-j}(x,k) \:,\label{sy1}
\end{eqnarray}
 where the coefficients 
(their number is infinite) fulfil the
homogeneity property $ A_{a-j}(x,tk)= t^{a-j}A_{a-j}(x,k)$, for $t>0$.
The number $a$ is called the order of $A$. 
Now, Wodzicki \cite{wod} proved that for two invertible,
 self-adjoint, elliptic, commuting, pseudodifferential operators 
on a smooth compact manifold without boundaries $M_D$:
 \begin{eqnarray} 
a(A,B)=\frac{\mbox{res}\left[
(\ln(A^bB^{-a}))^2 \right]}{2ab(a+b)}=a(B,A)
\:,\label{wod3}
\end{eqnarray}
where $a >0$ and $ b> 0$ are the orders
of $A$ and $B$, respectively.
Here the quantity $\mbox{res}(A)$ is the Wodzicki non-commutative
residue. It can be computed easily using the homogeneous component
$A_{-D}(x,k)$ of order $-D$ of the complete symbol, 
\begin{eqnarray}
\mbox{res}(A)=\int_{M_D}\frac{dx}{(2\pi)^{D}}\int_{|k|=1}A_{-D}(x,k)dk
\:.\label{wod2}
\end{eqnarray}

All this can be applied to our operators. As an example:
\beq
A(x,k)_{K_{\pm}}\!=\aq \ln \at k^2+m^2-e^2\mu^2+ i2e\mu  k_\tau
  \ct \cp \!\!-\!\!\ap \ln \at k^2+m^2-e^2\mu^2- i2e\mu  k_\tau \ct \cq^2
\:.\label{903}
\enq
Simply expanding (\ref{903}) and performing the above integration (\ref{wod2}) we obtain
the non-commutative residue. Remembering (\ref{wod3}) and that the
order of our operators is 2, we have the related multiplicative anomaly as
\beq
a_4(K_+,K_-)=\frac{\be V}{8\pi^2} \aq
e^2\mu^2( m^2-\frac{e^2\mu^2}{3}) \cq
\:.\label{ex1}
\enq
The same can be done for $L_{\pm}$, obtaining another expression
for $a_4(L_+,L_-)$.

Finally, including this two results in (\ref{441}) and (\ref{44})
respectively, we obtain
\beq
\ln Z_{\be}(K_+,K_-)=\ln Z_{\be}(L_+,L_-)&=&\frac{\be V }{32\pi^2}\aq
m^4 (\ln \fr{m^2}{M^2}-3/2) \cq+S(\be,\mu)\nn \\
                        &-&\frac{\be V}{16\pi^2} \aq
e^2\mu^2( m^2-\frac{e^2\mu^2}{3}) \cq\: ,\label{441bis}
\enq
and the logarithm of the partition function turns to be the same for
the two different approaches.
Although consistent now, our result is remarkably different from the
one in the literature where the multiplicative anomaly was
disregarded. The physical relevance of this additional term
will be discussed in the next section.

More generally, this term can be easily computed for any space-time
dimension $D$ and turns out to be always vanishing for odd $D$ 
\cite{evz,efvz}. 
It has also been computed for the self-interacting field
\cite{bbd,bd}, but there
many difficulties arise when dealing with the regularized determinants
of the complicated operators involved \cite{efvz}.

\section{PHYSICAL RELEVANCE}

To investigate the physical relevance of the multiplicative anomaly
the crucial quantity is the effective potential in presence of external
sources, which can be expressed as a function of the charge
density $\rho=\frac{1}{\be V}\frac{\partial \ln
Z_{\be}(\mu,J_i)}{\partial \mu}=\frac{<Q>}{V}$  
and the mean field $x^2=\Phi^2 $ as
\beq
F(\be,\rho,x)=-\frac{1}{\be V}\ln Z_{\be}(\mu)
+\frac{\mu}{\be V} \frac{\partial \ln Z_{\be}(\mu) }{\partial
\mu}+\frac{1}{2}(m^2+e^2\mu^2)x^2 \:,
\label{v1}
\enq
\begin{eqnarray} 
\rho=\frac{1}{\be V}\frac{\partial \ln
Z_{\be}(\mu)}{\partial \mu}+ e^2\mu x^2 \:,\label{v2}
\end{eqnarray}
where the later is an implicit expression for the chemical potential
as a function of $\rho$.
 
The physical states correspond to the minima of the effective
potential, located in $\frac{\partial F}{\partial x}= x
(m^2-e^2\mu^2)=0$. We find therefore: 1) an unbroken phase, $x=0$,
$e\mu <m$;
2) a symmetry breaking solution, $x\neq 0$, $e\mu=\pm m$,
giving the relativistic Bose-Einstein condensation.
For our system, explicitly, the unbroken and broken phase are respectively
\beq
{\cal F}_\be= min F &=&{\cal E}_V-\frac{1}{\be V} S(\be,\mu)+\mu \rho +\frac{1}{16\pi^2} \aq
e^2\mu^2( m^2-\frac{e^2\mu^2}{3}) \cq, \\
\rho &=&-\frac{1}{\be V}\frac{\partial S(\be,\mu)}{\partial
\mu}-\frac{e}{8\pi^2} \aq
e\mu( m^2-\frac{2 e^2\mu^2}{3}) \cq\: , 
\label{sbp111}
\enq
 where ${\cal E}_V$ is the vacuum contribution, and
\beq
{\cal F}_\be &=&{\cal E}_V-\frac{1}{\be V} S(\be,e\mu=m)+\frac{m}{e} \rho +\frac{1}{8\pi^2}\frac{m^4}{3} , \\
\rho &=&-\frac{1}{\be V}\frac{\partial S(\be,\mu)}{\partial
\mu}|_{e\mu=m}-\frac{e}{8\pi^2}\frac{m^3}{3}
+e m x^2
\:.\label{sbp1}
\enq
It is possible to see, under a detailed generic $D$
analysis, that these expressions, as they are, give some
inconsistencies in the broken phase \cite{efvz}.
We have to remember, though, that we worked until now with regularized but
``unrenormalized'' charge density. Since it appears in the partition
function multiplied by $\mu$, any ambiguity in it will correspond to
an uncertainty in the free energy density of the kind $\mu K$. This $K$ has to be fixed following physical requirements. A very reasonable
one is that the symmetry is unbroken at $T=0$, $\rho=0$. For $D=4$, $K$
will be $K=-\frac{e m^3}{24\pi^2}$. For $D=4$ only, this choice also removes the
multiplicative anomaly contribution to the charge density, so that the anomaly does not alter the broken phase 
in any aspect and we get
 \begin{eqnarray}
{\cal F}_\beta &=&{\cal E}_V-\frac{1}{\beta V}
S(\beta,m)+\frac{m}{e}  \rho^R, \\ e x^2&=&
\left. \frac{1}{m}\left( \rho^R+\frac{1}{\beta V}\frac{\partial S(\beta,\mu)}{\partial \mu}\right|_{e\mu=m} \right)
\:.\label{sbp2}
\end{eqnarray}
Also the critical temperature ($x=0$, $e\mu=m$) remains unchanged with
this renormalization.
Is different the unbroken phase, where the anomalous term remains:
\begin{eqnarray} {\cal F}_\beta &=&{\cal E}_V-\frac{1}{\beta V} S(\beta,\mu)+\mu \rho^R -\frac{\mu e m^3}{12 \pi^2}+\frac{1}{8
\pi^2}\left( e^2\mu^2m^2-\frac{1}{3}e^4\mu^4 \right), \\
\rho^R&=&-\frac{1}{\beta V}\frac{\partial S(\beta,\mu)}{\partial
\mu} -\frac{1}{8 \pi^2}\left( e^2\mu m^2-\frac{2}{3}e^4\mu^3 \right)
+\frac{em^3}{24\pi^2}\:. \label{sspp}
\end{eqnarray}
We should now observe that the anomalous contribution to the free
energy is non leading at ultra relativistic temperatures $T > m$,
 since the thermal terms go as $T^4$. On the other
hand, it does not even contribute to the low temperature limit,
corresponding to the broken phase, so that it could give relevant
corrections only in a intermediate range $T\simeq m $. 
Notice, finally, that the anomalous term is vanishing as
 $e \to 0$, and the correct expression of the free energy
density for the uncharged boson gas is recovered.

\section{GENERAL CONSIDERATIONS}

This analysed is just one of the many possible physical systems were
the multiplicative anomaly could play a role \cite{nuovozerbini}.
 The first question is therefore
if this additional terms will always have physical relevance \cite{dowker,rdowker}.
This will in general depend on the system. 
As an example, for the above case the anomaly is
vanishing for any odd dimension. In other cases \cite{evz} the anomaly could be
 simply non physical, as it could be reabsorbed in 
the renormalization procedure. Work is currently in progress on
other systems, including fermionic ones. It is not difficult to see
that for a single free fermionic field the anomaly is always
vanishing, too. On the other side, it could play a role for neutrino
mixing, in relation with recent results regarding inequivalent
representations of the vacuum \cite{mixing} and, in general, any time when there is a
possible mixing or rotation in the functional space of the fields \cite{rdowker}.
This needs further investigation due to the deep
connection between the multiplicative anomaly and the functional measure, which
goes to the roots of the definition of the functional integral itself.

The other relevant question is, of course, if the anomaly is
regularization dependent \cite{evans}. Zeta function regularization is just 
one example of a wider class of regularizations called: ``generalized
proper-time regularizations'' \cite{sch,bal}, for which we showed the
anomaly to be present \cite{bcvz,efvz}. 
This topic is also under further investigations and created a vivid
debate lately. Here too, the answers are probably to be found in a
proper and consistent definition of the ill-defined functional 
determinant itself, where this regularization approach has,
 up to now, proved to be rigorous and coherent \cite{revans}.

{\em Note added in proof:} After my talk, a work \cite{mcktom} by
McKenzie-Smith and Toms appeared.
There, the relevance of considering the multiplicative
anomaly within a functional integral approach is recognized although 
they do not agree on its physical relevance
for the relativistic charged bosonic field.

\end{document}